\documentclass[12pt]{article}
\usepackage[sort&compress,square,comma,numbers]{natbib}
\usepackage{amsmath,amssymb,graphicx,slashed}

\usepackage[
	colorlinks=true,
	citecolor=black,
	linkcolor=black,
	urlcolor=blue,
	hypertexnames=false]{hyperref}

\newcommand{\arXiv}[2]{\href{http://arxiv.org/pdf/#1}{{\tt #2/#1}}}
\newcommand{\arXivold}[1]{\href{http://arxiv.org/pdf/#1}{{\tt #1}}}

\newcommand{\beq}{\begin{eqnarray}}
\newcommand{\eeq}{\end{eqnarray}}

\begin{document}
\begin{center} 
{\huge \bf Dark Monopoles  \\ \vspace*{0.25cm} 
and $SL(2,\mathbb{Z})$ Duality \vspace*{0.5cm}} 
\end{center}

\begin{center} 

{\bf John Terning} and {\bf Christopher B. Verhaaren} \\
\end{center}
\vskip 8pt
\begin{center} 
{\it Center for Quantum Mathematics and Physics (QMAP)\\Department of Physics, University of California, Davis, CA 95616} 
\end{center}

\vspace*{0.1cm}
\begin{center} 
{\tt 
 \href{mailto:jterning@gmail.com}{jterning@gmail.com}\,
 \href{mailto:cbverhaaren@ucdavis.edu}{cbverhaaren@ucdavis.edu}}

\end{center}

\centerline{\large\bf Abstract}
\begin{quote}
We explore kinetic mixing between two Abelian gauge theories that have both electric and magnetic charges. When one of the photons becomes massive,
novel effects arise in the low-energy effective theory, including the failure of Dirac charge quantization as particles from one sector obtain parametrically small couplings to the photon of the other. We maintain a manifest $SL(2,\mathbb{Z})$ duality throughout our analysis, which is the diagonal subgroup of the dualities of the two un-mixed gauge theories.
\end{quote}


\section{Introduction}
Recently bounds have been placed \cite{Hook:2017vyc} on dark matter models containing a massive dark photon and dark magnetic monopoles when there is kinetic mixing \cite{Holdom:1985ag} of the dark and ordinary photons \cite{Brummer:2009cs,Bruemmer:2009ky,Sanchez:2011mf}. The bounds crucially rely on the effective coupling of the ordinary photon to the dark monopole, and on the fact
that this coupling, in general, violates Dirac-Schwinger-Zwanziger charge quantization \cite{Dirac,Schwinger,Zwanziger:1968rs}. While we believe the bounds in \cite{Hook:2017vyc} are essentially correct, the theoretical derivation of the coupling of the ordinary photon to dark magnetic charges is technically incorrect when there are both electric \emph{and} magnetic charges in the dark sector. The general case, with both types of charges, is more subtle because it is impossible to write a local, Lorentz invariant action with both electric and magnetic charges \cite{Dirac2,Hagen,Zwanziger:1971}. Thus, the low-energy theory with a single photon cannot have a local, Lorentz invariant Lagrangian.

Zwanziger \cite{Zwanziger:1971} showed that it is possible to write a local Lagrangian for electric and magnetic charges at the cost of requiring two gauge potentials (one with local couplings to electric charges and one with local couplings to magnetic charges) and the loss of manifest Lorentz invariance. Having two gauge potentials appears to double the number of propagating photon polarizations, however Zwanziger's kinetic terms project out two polarizations on-shell by introducing a four-vector $n^\mu$ that breaks manifest Lorentz invariance. In certain gauges this four vector can be identified with the direction of the Dirac string, so it is plausible that it is a gauge artifact, and formal proofs have been given \cite{Brandt:1978} to show that physical observables are $n^\mu$ independent. 

It was later seen \cite{Csaki} that this type of Lagrangian realizes $SL(2,\mathbb{Z})$ duality \cite{Cardy1,Vafa,Witten2,Lozano,Strominger:2015bla} as a local field redefinition. Furthermore, the $SL(2,\mathbb{Z})$ duality of the low-energy effective theory is maintained order-by-order in the loop expansion \cite{Laperashvili:1999pu,Colwell:2015wna}. In the absence of kinetic mixing, each $U(1)$ sector has its own $SL(2,\mathbb{Z})$ duality, but when kinetic mixing occurs, typically at loop level through particles charged under both sectors~\cite{Holdom:1985ag}, the duality is reduced to a diagonal $SL(2,\mathbb{Z})$ subgroup.

In this paper we provide a firm theoretical derivation for the results of \cite{Hook:2017vyc} while maintaining manifest $SL(2,\mathbb{Z})$ duality. We review the (not widely known) two potential formalism, $SL(2,\mathbb{Z})$ duality, and the renormalization of such theories. 
We then introduce kinetic mixing between the ordinary photon and a dark sector photon with both electric and magnetic charges. 
Using the two-potential formalism in both the dark and visible sectors, we show how the diagonal gauge sector's $SL(2,\mathbb{Z})$ is a subgroup of two $SL(2,\mathbb{Z})$ dualities. Finally, we consider the cases of dark electrically and magnetically  charged condensates, which lead to a dark photon mass and confinement of dark magnetic or electric charges. We clarify how these confining scenarios are directly connected to parametrically small (charge quantization violating) couplings of dark sector states to the visible photon.

\section{An $SL(2,\mathbb{Z})$ Covariant Lagrangian}
We begin with a single $U(1)$ gauge theory having both electric and magnetic charges. In order to have a local Lagrangian, we use Zwanziger's two potential formulation \cite{Zwanziger:1971} of QED.\footnote{ A special case of this formulation was independently rediscovered by Schwarz and Sen \cite{Schwarz:1993vs}; they referred to their action as duality symmetric since under $SL(2,\mathbb{Z})$ duality transformations the gauge potentials undergo canonical transformations resulting in a new action that only differs in its coupling constant.}
The $A_\mu$ and $B_\mu$ gauge potentials have local couplings to electric and magnetic currents respectively. For electric and magnetic charges satisfying the Dirac-Schwinger-Zwanziger \cite{Dirac,Schwinger,Zwanziger:1968rs} charge quantization condition, $A_\mu$ has coupling strength 
$e$, while $B_\mu$ has coupling strength $4\pi/e$.
In the differential form notation of~\cite{Csaki}, the Lagrangian is:
\begin{align}
\mathcal{L}=&-\text{Im}\left\{\frac{\tau}{8\pi n^2}\left[n\cdot\partial \wedge\left( A+iB\right) \right]\cdot\left[ n\cdot\partial \wedge\left( A-iB\right)\right] \right\}\nonumber\\
&- \text{Re}\left\{\frac{\tau}{8\pi n^2}\left[n\cdot\partial \wedge\left( A+iB\right) \right]\cdot\left[ n\cdot{}^\ast\partial \wedge\left( A-iB\right)\right] \right\}\nonumber\\
&-\text{Re}\left[ (A-iB)\cdot(J+\tau K)\right],\label{e.holoLag}
\end{align}
where $J$ and $K$ are the electric and magnetic currents respectively and the holomorphic coupling is
\beq
\tau=\frac{\theta}{2\pi}+\frac{4\pi i}{e^2}~.
\eeq 
The Lagrangian in \eqref{e.holoLag} clearly depends on the constant four-vector $n^\mu$, apparently violating Lorentz invariance. While it has been formally argued~\cite{Brandt:1978} that gauge invariant observables are independent of $n^\mu$, this is typically difficult to see operationally. In special cases, such as the soft-photon limit~\cite{Terning:2018udc}, when an all-orders calculation can be completed the dependence on $n^\mu$ vanishes from observables when charge quantization is satisfied.

It is convenient to unpack Eq.~\eqref{e.holoLag} using indexed fields. In doing so, we use the definition
\beq
F^X_{\mu\nu}\equiv\partial_\mu X_\nu-\partial_\nu X_\mu~,
\eeq
to obtain
\begin{align}
\mathcal{L}=&-\frac{n^\alpha n^\mu}{8\pi n^2} g^{\beta\nu}\left[\frac{4\pi}{e^2}\left(F^A_{\alpha\beta}F^A_{\mu\nu}+ F^B_{\alpha\beta}F^B_{\mu\nu}\right)\right]\nonumber\\
&-\frac{n^\alpha n_\mu}{16\pi n^2}\varepsilon^{\mu\nu\gamma\delta}\left[ \frac{\theta}{2\pi}\left(F^A_{\alpha\nu}F^A_{\gamma\delta}+F^B_{\alpha\nu}F^B_{\gamma\delta} \right)-\frac{4\pi}{e^2}\left( F^B_{\alpha\nu}F^A_{\gamma\delta}-F^A_{\alpha\nu}F^B_{\gamma\delta} \right)\right]\nonumber\\
&-J_\mu A^\mu-\frac{4\pi}{e^2}K_\mu B^\mu-\frac{\theta}{2\pi}A_\mu K^\mu.\label{e.Lagtheta}
\end{align}
The equations of motion (i.e. the Maxwell Equations)  for $A_\mu$ and $B_\mu$ are
\beq
\frac{\text{Im}(\tau)}{4\pi}\,\partial_\nu\left(F^{\mu\nu}+i{}^\ast\! F^{\mu\nu} \right)=J^\mu+\tau K^\mu,
\label{Maxwell}
\eeq
where
\begin{align}
F_{\mu\nu}=&\frac{n^\alpha}{n^2}\left(n_\mu F^A_{\alpha\nu}-n_\nu F^A_{\alpha\mu}-\varepsilon_{\mu\nu\alpha}^{\phantom{\mu\nu\alpha}\beta}n^\gamma F^B_{\gamma\beta} \right)~,\label{e.Fdef}\\
{}^\ast\! F_{\mu\nu}=&\frac{n^\alpha}{n^2}\left(n_\mu F^B_{\alpha\nu}-n_\nu F^B_{\alpha\mu}+\varepsilon_{\mu\nu\alpha}^{\phantom{\mu\nu\alpha}\beta}n^\gamma F^A_{\gamma\beta} \right)~.
\label{e.Fstardef}
\end{align}

Note that $\theta$ only appears in the Maxwell equation through the coupling of the magnetic current to the electric gauge potential $A_\mu$, which is just the Witten effect \cite{Witten}. Both the field strength and the propagator are independent of $\theta$. Indeed, one can check Zwanziger's derivation of the propagator~\cite{Zwanziger:1971}, without the $\theta$ dependent kinetic term,
\beq
\mathcal{L}_\theta= -\frac{n^\alpha n_\mu}{16\pi n^2}\varepsilon^{\mu\nu\gamma\delta}\ \frac{\theta}{2\pi}\left(F^A_{\alpha\nu}F^A_{\gamma\delta}+F^B_{\alpha\nu}F^B_{\gamma\delta} \right),
\label{thetaterm}
 \eeq
 to see that he has the same equations of motion with $\tau$ purely imaginary. Thus, $\mathcal{L}_\theta$ can be omitted from the Lagrangian without changing the dynamics of the theory. By dropping this term can rewrite the Lagrangian in Eq.~\eqref{e.Lagtheta} as
\begin{align}
\mathcal{L}=&-\frac{n^\alpha n^\mu\,\text{Im}(\tau)}{8\pi n^2}\left[g^{\beta\nu}\left(F^A_{\alpha\beta}F^A_{\mu\nu}+F^B_{\alpha\beta}F^B_{\mu\nu} \right)-\frac12\varepsilon_{\mu}^{\phantom{\mu}\nu\gamma\delta}\left(F^B_{\alpha\nu}F^A_{\gamma\delta}-F^A_{\alpha\nu}F^B_{\gamma\delta} \right) \right]\nonumber\\
&-\text{Re}\left[ \left( A-iB\right)\cdot\left( J+\tau K\right)\right]~.
\label{ABLagrange}
\end{align}
There are two other useful ways to write this Lagrangian:
\beq
\mathcal{L}&=&-\frac{\text{Im}(\tau)}{16\pi}\left[F^{\mu\nu}F^A_{\mu\nu}+{}^\ast\! F^{\mu\nu}F^B_{\mu\nu}\right]-\text{Re}\left[ \left( A-iB\right)\cdot\left( J+\tau K\right)\right]
\label{easy}\\
&=&-\frac{\text{Im}(\tau)}{32\pi}\left[F_+^{\mu\nu}\left(F^A_{\mu\nu}-iF^B_{\mu\nu} \right)+F_-^{\mu\nu}\left(F^A_{\mu\nu}+iF^B_{\mu\nu} \right)\right] 
\label{manifest}\\
&&-\text{Re}\left[ \left( A-iB\right)\cdot\left( J+\tau K\right)\right]~. \nonumber
\eeq
The version in Eq.~\eqref{easy} makes the derivation of the equations of motion particularly easy, while Eq.~\eqref{manifest} makes the $SL(2,\mathbb{Z})$ covariance manifest. As we will see, this is the correct low-energy effective theory for a topological monopole \cite{'tHooft:1974qc} for energies much smaller than the inverse size of the monopole core. The only further requirement is that the spectrum of charged particles is free of electric, magnetic, and mixed electric/magnetic anomalies \cite{Csaki}.

Under an $SL(2,\mathbb{Z})$ duality transformation \cite{Cardy1,Vafa,Witten2,Lozano} the currents are mapped to
\beq
J^\mu\to bK'^\mu+dJ'^\mu, \ \ \ \ K^\mu\to aK'^\mu+cJ'^\mu.
\eeq
where $a,b,c,d$ are integers with $ad-bc=1$. 
The gauge fields transform \cite{Csaki} as
\beq
A_\mu+iB_\mu &\to& \frac{1}{c\tau^\ast+d}\left( A'_\mu+i B'_\mu\right)~, \\
A_\mu-iB_\mu &\to& \frac{1}{c\tau+d}\left( A'_\mu-i B'_\mu\right).
\eeq
From Eqs.~\eqref{e.Fdef} and \eqref{e.Fstardef} we can separate the field strength into irreducible representations of $SO(4)\sim SU(2)\times SU(2)$. Specifically, the field strength splits into  $(1,0)$ and $(0,1)$ representations, which are given by:
\beq
F_{\pm \mu\nu}&\equiv& F_{\mu\nu}\pm i\,{}^\ast\! F_{\mu\nu}\\
&=&\frac{n^\alpha}{n^2}\left[n_\mu\left(F^A_{\alpha\nu}\pm iF^B_{\alpha\nu} \right)-n_\nu\left(F^A_{\alpha\mu}\pm iF^B_{\alpha\mu} \right)\pm i\varepsilon_{\mu\nu\alpha}^{\phantom{\mu\nu\alpha}\beta}n^\gamma\left(F^A_{\gamma\beta}\pm i F^B_{\gamma\beta} \right) \right]~.\nonumber
\eeq
It is easy to check that
\beq
n^\mu F_{\pm\mu\nu}=n^\alpha\left(F^A_{\alpha\nu}\pm iF^B_{\alpha\nu} \right)=\left[n\cdot\partial\wedge(A\pm iB)\right]_\nu~.
\eeq
This means that the field strengths transform as
\beq
F_{\mu\nu}+ i\,{}^\ast\! F_{\mu\nu} &\to& \frac{1}{c\tau^\ast+d}\left(F'_{\mu\nu}+ i\,{}^\ast\! F'_{\mu\nu} \right)~, \\
F_{\mu\nu}- i\,{}^\ast\! F_{\mu\nu} &\to& \frac{1}{c\tau+d}\left(F'_{\mu\nu}- i\,{}^\ast\! F'_{\mu\nu}\right)~.
\eeq

The transformation of the interaction term,
\beq   
{\mathcal L}_{\rm int} =-\text{Re}\left[ \left( A-iB\right)\cdot\left( J+\tau K\right)\right]~,
\eeq
under $SL(2,\mathbb{Z})$ is
\beq
{\mathcal L}_{\rm int} &\to& -\text{Re}\left\{  \frac{A'_\mu-i B'_\mu}{c\tau+d}\left[ b K'^\mu+d J'^\mu+\tau \left( a K'^\mu+c J'^\mu \right)\right]\right\} \\
&\to& -\text{Re}\left[  \left(A'_\mu-i B'_\mu\right)\left( J'^\mu+\frac{a\tau+b}{c\tau+d} K'^\mu\right)\right] ,
\eeq
so that after the duality transformation the holomorphic coupling $\tau$ is replaced by 
\beq
\tau'=\frac{a\tau+b}{c\tau+d}~.
\eeq
This, in turn, implies a simple transformation law for the imaginary part of $\tau$: 
\beq
\text{Im}(\tau')=\frac{\text{Im}(\tau)}{\left|c\tau+d \right|^2}.
\eeq
This agrees exactly with the transformation of the kinetic terms,
\beq
{\mathcal L}_{\rm kin}=-\frac{\text{Im}(\tau)}{32\pi}\left[F_+^{\mu\nu}\left(F^A_{\mu\nu}-iF^B_{\mu\nu} \right)+F_-^{\mu\nu}\left(F^A_{\mu\nu}+iF^B_{\mu\nu} \right)\right] ~,
\eeq
which become
\beq
{\mathcal L}_{\rm kin} &\to& \frac{{\mathcal L}_{\rm kin} }{(c \tau +d)(c \tau^\ast+d)}~,
\eeq
Finally, note that the Maxwell Equations (\ref{Maxwell}) become
\beq
\frac{\text{Im}(\tau')}{4\pi}\partial_\nu\left(F'^{\mu\nu}+i{}^\ast\! F'^{\mu\nu} \right)=J'^\mu+\tau' K'^\mu~,
\eeq
which confirms that the dual coupling is indeed $\tau'$.

\section{Renormalization}
It is convenient when discussing the renormalization of the Zwanziger Lagrangian to rescale the fields so that the running coupling only appears in the kinetic terms. This is achieved by 
\beq
B_\mu \to e^2 B_\mu~.
\eeq
After this redefinition the Lagrangian becomes
\beq
\mathcal{L}&=&-\frac{n^\alpha n^\mu}{2 n^2}\left[g^{\beta\nu}\left(\frac{1}{e^2}F^A_{\alpha\beta}F^A_{\mu\nu}+e^2 F^B_{\alpha\beta}F^B_{\mu\nu} \right)-\frac12\varepsilon_{\mu}^{\phantom{\mu}\nu\gamma\delta}\left(F^B_{\alpha\nu}F^A_{\gamma\delta}-F^A_{\alpha\nu}F^B_{\gamma\delta} \right) \right]\nonumber\\
&&-J_\mu A^\mu-4\pi\,K_\mu B^\mu-\frac{\theta}{2\pi}A_\mu K^\mu.
\eeq
Integrating out particles with electric and/or magnetic charges results in vacuum polarization corrections. Explicit calculations \cite{Laperashvili:1999pu} reveal that the one-loop correction to the effective Lagrangian, renormalized at the scale $\mu$, from either electrically or magnetically charged particles is 
\beq
\mathcal{L}_{\rm one-loop}&=&\frac{n^\alpha n^\mu}{ n^2}g^{\beta\nu}\, \Pi(\mu)\left(F^A_{\alpha\beta}F^A_{\mu\nu}-F^B_{\alpha\beta}F^B_{\mu\nu} \right)=\frac{1}{2}\, \Pi(\mu)F_{\mu\nu}F^{\mu\nu}~,
\label{one-loop}
\eeq
with electric and magnetic charges contributing with opposite signs to $\Pi(\mu)$. Note that only one of $e$ or $b=4\pi/e$ can be small for the calculation, so one must consider only electric charges or only magnetic charges in the calculation, but either type of charge renormalizes both $A$ and $B$ terms. The analysis of refs. \cite{Argyres:1995jj,Csaki}, using $SL(2,\mathbb{Z})$ transformations, shows that particles with both electric and magnetic charges (i.e. dyons) also renormalize
the $\theta$ term in Eq. \eqref{thetaterm}. 

The Lagrangian can be canonically normalized by rescaling
\beq
A_\mu \to \sqrt{Z_3^A(\mu)}\,e\,A_\mu~,\quad B_\mu \to \sqrt{Z_3^B(\mu)}\,B_\mu/e~,
\eeq
so that the renormalized electric and magnetic couplings are, respectively,
\beq
e(\mu)= \sqrt{Z_3^A(\mu)} \,e~, \quad
b(\mu)=\sqrt{Z_3^B(\mu)}\, \frac{4 \pi}{e}~.
\label{ebrenorm}
\eeq
$SL(2,\mathbb{Z})$ duality, and the self-consistency of the Lagrangian, requires that
\beq
Z_3^A(\mu)=1/Z_3^B(\mu)~,
\label{SL2Zrenorm}
\eeq
which, from (\ref{one-loop}), is certainly true at one-loop:
\beq
Z_3^A(\mu)=1/Z_3^B(\mu)=1-\Pi(\mu)~.
\label{SL2Zrenormloop}
\eeq
Note, however, that at higher loop order Eq.~\eqref{SL2Zrenorm} requires a more complicated relation between the corrections to the $A_\mu$ and $B_\mu$ kinetic terms.

As Coleman \cite{Coleman:1982cx} pointed out, Eq. (\ref{SL2Zrenorm}) is essential for understanding the renormalized Dirac-Schwinger-Zwanziger charge quantization condition. To understand why this is the case, suppose that we do not impose (\ref{SL2Zrenorm}), and that  $e(\mu)$ and $b(\mu)$ are independently renormalized. Dirac originally found the charge quantization condition \cite{Dirac} by imagining what would come to be known as an Aharonov-Bohm experiment \cite{Aharonov} that transports an electron with electric charge -1 around the Dirac string of a monopole with a half-integer magnetic charge $n/2$. Upon returning to the starting point there is a relative phase
\beq
- e(\mu) b(\mu) \frac{n}{2}= -\sqrt{Z_3^A(\mu)Z_3^B(\mu)} \,2 \pi n~.
\eeq
Coleman's point is that this phase must be a multiple of $2 \pi$ for any renormalization scale smaller than the inverse size of the monopole, i.e. it is a renormalization group invariant of the low-energy effective theory. Thus, Eq. (\ref{SL2Zrenorm})  must be true order-by-order in perturbation theory, and
\beq
b(\mu)= \frac{4 \pi}{e(\mu)}~,
\eeq
which is equivalent to Eq. (\ref{SL2Zrenorm}).

As pointed out in ref. \cite{Colwell:2015wna}, this entire renormalization discussion is completely consistent with the exact results of the Seiberg-Witten theory \cite{SeibergWitten}. In the region of the moduli space where the monopole is light, the low-energy electric coupling $e(\mu)$ is driven to large values as we run $\mu$ down to the monopole mass from higher energy scales. Hence, for a sufficiently small monopole mass, the monopole is weakly coupled to $B_\mu$, and since there are no light electrically charged particles, the entire  low-energy theory is weakly coupled.  Performing an $SL(2,\mathbb{Z})$ transformation which takes the light monopole to a light electron also takes the electric coupling to a dual electric coupling $e'(\mu)=4\pi/e(\mu)$.  Since $e'(\mu)$ is also small we see that the dual low-energy theory is also weakly coupled and the usual perturbative running drives the coupling $e'(\mu)$ to smaller values as we lower the renormalization scale toward the mass of the electrically charged particle. Thus, we see that electrically and magnetically charged particles drive the coupling $e(\mu)$ in opposite directions, as seen in refs. \cite{Laperashvili:1999pu,Csaki,Argyres:1995jj}.

\section{Mixing With A Dark Sector}
We now include a separate ``dark" sector, denoted by a subscript $D$, with a Lagrangian:
\begin{align}
\mathcal{L}_D=&-\frac{n^\alpha n^\mu}{2 n^2}\left[g^{\beta\nu}\left(F^A_{D\alpha\beta}F^A_{D\mu\nu}+F^B_{D\alpha\beta}F^B_{D\mu\nu} \right)-\frac12\varepsilon_{\mu}^{\phantom{\mu}\nu\gamma\delta}\left(F^B_{D\alpha\nu}F^A_{D\gamma\delta}-F^A_{D\alpha\nu}F^B_{D\gamma\delta} \right) \right]\nonumber\\
&-e_DJ_{D\mu} A_D^\mu-\frac{4\pi}{e_D}K_{D\mu} B_D^\mu~.
\end{align}
Note that for the remainder of the paper we work in canonical normalization. We assume the same form for the visible sector by taking $A_\mu,B_\mu\to eA_\mu, eB_\mu$ in Eq. (\ref{ABLagrange}). One might expect a distinct $n^\mu_D$ as well, but for simplicity we choose gauges in which these vectors are the same in each sector. We also set the CP violating $\theta$'s to zero, so that we can ignore the subtleties that arise from relating the $\theta$'s in each sector. The $\theta$ dependent electric charge of a monopole depends on the details of zero modes of electrically charged particles \cite{Callan:1982au,Harvey:1983tp,Niemi:1984dq}. This means the $\theta$ dependence of the theory is partially determined by the UV completion of the monopoles, and is thus not purely controlled by the low-energy effective theory, so we leave these details for future study.

Any heavy particles that are charged under both $A_\mu$ and $A_{D\mu}$  can lead to kinetic mixing \cite{Holdom:1985ag} between the two sectors. As described in the previous section, renormalization effects in the Zwanziger Lagrangian are somewhat subtle, but have been explored in detail at one-loop in ref. \cite{Laperashvili:1999pu}. At one-loop, analogous to  Eq.~\eqref{one-loop}, the following kinetic mixing term is generated:
\beq
\mathcal{L}_\epsilon=\epsilon ee_D\frac{n^\alpha n^\mu}{n^2}g^{\beta\nu}\left(F^A_{D\alpha\beta}F^A_{\mu\nu}-F^B_{D\alpha\beta}F^B_{\mu\nu} \right)=\frac{\epsilon e e_D}{2}F_{\mu\nu}F_D^{\mu\nu},\label{e.kinMix}
\eeq
where we have kept the gauge couplings factors distinct from $\epsilon$. We emphasize here that this simple form holds only at one-loop. Because the renormalization of $e(\mu)$ and $b(\mu)$ have an inverse relationship, as seen in Eqs.~\eqref{ebrenorm} and \eqref{SL2Zrenorm},  the charge quantization condition at higher loop order will require a more complicated structure. Therefore, in what follows we work to linear order in $\epsilon$.

We should also pause to consider the $SL(2,\mathbb{Z})$ behavior of the mixing term. By rewriting Eq.~\eqref{e.kinMix} as
\beq
\mathcal{L}_\epsilon=\epsilon ee_D\frac{n^\alpha n^\mu}{n^2}g^{\beta\nu}\left(F^A_{D\alpha\beta}+iF^B_{D\alpha\beta} \right)\left(F^A_{\mu\nu}+iF^B_{\mu\nu} \right),
\eeq
it is clear that the plus representation, $(A_D+iB_D)$, of the dark sector must transform like the minus representation, $(A-iB)$, of the visible sector. In this way the diagonal subgroup $SL(2,\mathbb{Z})$ is preserved by the kinetic mixing.

Now that we know the form of the kinetic mixing, we can transform to a diagonal basis, denoted by $\overline{V}_\mu$, where $V_\mu$ is any gauge potential. The diagonalization is achieved  by
\begin{align}
\left( \begin{array}{c}
A_\mu\\
A_{D\mu}
\end{array}\right)=&\left(\begin{array}{cc}
\cos\phi+\epsilon ee_D\sin\phi & -\sin\phi+\epsilon ee_D\cos\phi \\
\sin\phi& \cos\phi
\end{array} \right)\left( \begin{array}{c}
\overline{A}_\mu\\
\overline{A}_{D\mu}
\end{array}\right),\\
\left( \begin{array}{c}
B_\mu\\
B_{D\mu}
\end{array}\right)=&\left(\begin{array}{cc}
\cos\phi & -\sin\phi \\
\sin\phi-\epsilon ee_D\cos\phi & \cos\phi+\epsilon ee_D\sin\phi
\end{array} \right)\left( \begin{array}{c}
\overline{B}_\mu\\
\overline{B}_{D\mu}
\end{array}\right).
\end{align}
The angle $\phi$ parametrizes a family of arbitrary $SO(2)$ rotations of the fields which preserve the diagonal kinetic terms. Importantly, the transformation matrices for $A$ and $B$ are not the same. Indeed, one must be the inverse of the transpose of the other so that the $AB$ mixing terms in Eq. (\ref{ABLagrange}) map to $\overline{A}\,\overline{B}$ mixing terms in the diagonal basis. 
In this basis the diagonal currents are given by
\begin{align}
\left( \begin{array}{c}
e\overline{J}_\mu\\
e_D\overline{J}_{D\mu}
\end{array}\right)=&\left(\begin{array}{cc}
\cos\phi+\epsilon ee_D\sin\phi& \sin\phi \\
-\sin\phi+\epsilon ee_D\cos\phi & \cos\phi
\end{array} \right)
\left( \begin{array}{c}
eJ_\mu\\
e_DJ_{D\mu}
\end{array}\right),\label{e.Jmix}\\
\left( \begin{array}{c}
\overline{K}_\mu/e\\
\overline{K}_{D\mu}/e_D
\end{array}\right)=&\left(\begin{array}{cc}
\cos\phi &  \sin\phi-\epsilon ee_D\cos\phi \\
 -\sin\phi & \cos\phi+\epsilon ee_D\sin\phi
\end{array} \right)
\left( \begin{array}{c}
K_\mu/e\\
K_{D\mu}/e_D
\end{array}\right).\label{e.Kmix}
\end{align}
This agrees with the analysis of~\cite{Hook:2017vyc}, leading to a $\epsilon$ magnetic charge from $K_D$ in $\overline{K}$ when $\phi=0$. That the $K$ and $J$ currents transform differently is an immediate consequence of the different transformations of $A$ and $B$. Without a physical reason to select a particular $\phi$, one can pick whatever is convenient for the calculation at hand. In particular, we could choose $\tan\phi=e\,e_D\,\epsilon$, where the coupling of $\epsilon$ magnetic charges to $\overline{B}_{\mu}$ vanishes. This illustrates that these $\epsilon$ magnetic charges are not physical without breaking the $SO(2)$ symmetry of the kinetic terms. This is done in the next section by giving a mass to the dark photon. 

\subsection{Charge Quantization}
Another physical manifestation of the charge quantization condition is associated with the angular momentum \cite{Goldhaber:1965cxe} carried by the gauge field accompanying an electric and a magnetic charge. J.J. Thomson showed \cite{Thomson} that the electromagnetic field of a static electric charge $q$ and a static magnetic charge $g$ carries an angular momentum
\beq
\vec{L}=qg\,\hat{r},
\eeq 
where $\hat{r}$ is the unit vector pointing from the electric charge to the magnetic charge. Quantum mechanics requires that angular momentum comes in half-integer units, which implies that
\beq
qg=\frac{m}{2}~,
\label{angularm}
\eeq
for some integer $m$. 

In the absence of kinetic mixing, the condition \eqref{angularm} is satisfied for each $U(1)$ sector individually. In particular, the dark sector electric charge $q_D$ and magnetic charge $g_D$ satisfy
\beq
q_D g_D=\frac{m_D}{2}~.
\label{angularmD}
\eeq
For non-zero mixing Eqs.~\eqref{e.Jmix} and \eqref{e.Kmix} imply
\begin{align}
\overline{q}\,\overline{g}=&qg\left(\cos^2\phi+ee_D\frac{\epsilon\sin 2\phi}{2} \right)+qg_D\left(\frac{\sin 2\phi}{2}-\epsilon ee_D\cos 2\phi\right)\nonumber\\
&+q_Dg ee_D\frac{\epsilon\sin 2\phi}{2} +q_Dg_D\left(\sin^2\phi-ee_D\frac{\epsilon\sin 2\phi}{2}  \right),\\
\overline{q}_D\overline{g}_D=&qg\left(\sin^2\phi-ee_D\frac{\epsilon\sin 2\phi}{2} \right)-qg_D\left(\frac{\sin 2\phi}{2}-ee_D\epsilon\cos2\phi\right)\nonumber\\
&-ee_Dq_Dg\frac{\epsilon\sin 2\phi}{2} +q_Dg_D\left(\cos^2\phi+ee_D\frac{\epsilon\sin 2\phi}{2}  \right),
\end{align}
which gives
\beq
\overline{q}\,\overline{g}+\overline{q}_D\overline{g}_D=qg+q_Dg_D=\frac{m+m_D}{2}~.
\eeq
Note, however, that when kinetic mixing occurs, $\overline{q}\,\overline{g}$ and $\overline{q}_D\overline{g}_D$ do not satisfy the quantization condition individually. Just as only a diagonal $SL(2,\mathbb{Z})$ is preserved in the mixed theory, so too there is only one global charge quantization shared between the two sectors. This makes perfect sense since the system of two particles has a single angular momentum and there is a single Aharonov-Bohm phase that could, in principle, be observed.

\section{Dark Confinement}
If we introduce an dark electrically charged scalar field and arrange its potential so that it has a VEV, then the dark photon will undergo a Higgs mechanism and become massive.  If the physical Higgs scalar is very heavy and has no other couplings to dark fields, we can simply add a mass term to our Lagrangian:
\beq
{\mathcal L}_{A_D\,{\rm mass}}=-\frac{m_{DA}^2}{2}A_{D\mu} A_D^\mu~.
\eeq
This means that dark magnetic charges coupled to $B_{D\mu}$ are confined \cite{'tHooftMandelstam}. A dark monopole-antimonopole pair is connected by a Nielsen-Olesen flux tube \cite{Nielsen:1973cs,Gubarev:1998ss} that behaves like a string with tension $\mathcal{O}(m_{DA}^{2})$. 

When the dark photon gets a mass, there is a preferred value of $\phi$: only $\sin\phi=0$ keeps the  $\overline{A}_\mu$ photon massless, as required by our unbroken $U(1)_\text{EM}$. In this case 
\begin{align}
\overline{J}_\mu=&J_\mu, &\overline{J}_{D\mu}=&J_{D\mu}+\epsilon e^2J_\mu,\nonumber\\
\overline{K}_\mu=&K_\mu -\epsilon e^2 K_{D\mu}, &\overline{K}_{D\mu}=&K_{D\mu}.
\end{align}
In short, the electric charges from the visible sector pick up an additional $\epsilon \,e^2\,e_D$ electric coupling to the massive dark photon while dark magnetic charges get an $\epsilon \,e\,4\pi$ coupling to the massless photon. This dark magnetic confinement is the scenario whose phenomenology was investigated in~\cite{Hook:2017vyc}.

If instead a dark magnetically charged scalar gets a VEV, the mass term
\begin{align}
{\mathcal L}_{B_D\,{\rm mass}}=&-\frac{m_{DB}^2}{2}B_{D\mu} B_D^\mu~,
\end{align}
results. In this case the dark electric charges are confined by electric flux tubes. In order that $\overline{B}$ remain massless, maintaining the unbroken $U(1)_\text{EM}$ in our sector, requires $\tan\phi=e\, e_D\,\epsilon.$ This gives the diagonal currents
\begin{align}
\overline{J}_\mu=&J_\mu+\epsilon e_D^2J_{D\mu}, &\overline{J}_{D\mu}=&J_{D\mu},\nonumber\\
\overline{K}_\mu=&K_\mu, &\overline{K}_{D\mu}=&K_{D\mu}-\epsilon e_D^2K_\mu.
\end{align}
In this case electrically charged dark particles get an $\epsilon \,e_D^2\, e$ electric coupling to the photon and any magnetically charged particles in our sector have an $\epsilon\, e_D\, 4\pi $ dark magnetic coupling. 


\section{Conclusion}
While the possibilities of particles with small electric charge under our photon have been discussed for some time~\cite{Holdom:1985ag}, detailed investigation of small magnetic charges has been more recent~\cite{Brummer:2009cs,Bruemmer:2009ky,Sanchez:2011mf}. This is at least partly due to the significant complication inherent in dealing with electric and magnetic charges simultaneously. The phenomenological analysis completed in \cite{Hook:2017vyc} investigated how such particles can be detected; however, up until now the theoretical derivation of these parametrically small magnetic charges has been lacking. 

In this work we have addressed this by using Zwanziger's local Lagrangian with both electric and magnetic charges. This formulation allows a clean analysis of the one-loop kinetic mixing between two otherwise separate $U(1)$ sectors. We have shown how this mixing can lead to milli-magnetically or milli-electrically charged particles. At the same time we have tracked the $SL(2,\mathbb{Z})$ duality enjoyed by both sectors and the related charge quantization condition preserved by the two sectors when mixed together. 

Spreading the charge quantization condition over two sectors is particularly useful for understanding the scattering of electric and magnetic charges. In a single sector, charge quantization implies that single photon exchange between an electric and magnetic charge goes like $e\cdot 4\pi/e$. Clearly this does not provide a good perturbative expansion. Consequently, there is no reason to be surprised by 
Weinberg's \cite{Weinberg:1965rz} demonstration that in this case single photon exchange is not Lorentz invariant. However, kinetic mixing combined with dark magnetic confinement, can lead to electric-magnetic scattering that goes like $\,e\cdot \epsilon \, e\,4\pi$, which can be small for $\epsilon\ll 1$. Therefore, this framework provides a useful laboratory for the perturbative understanding of the scattering of electric and magnetic charges~\cite{Terning:2018udc}.

It is interesting to note that Nambu has shown \cite{Nambu:1977ag} that an analogous situation arises in the standard model. Electroweak symmetry breaking via the Higgs doublet VEV leads to confined monopole---anti-monopole pairs. The mass mixing of the gauge bosons causes the massive $Z$ to carry part of the magnetic charge through a flux-tube, while a smaller fraction of the magnetic charge couples to the massless photon. Thus, the analysis presented here can be employed for analyzing these standard model partially-charged monopoles.

\appendix
\section*{Acknowledgments}
JT thanks Neil Weiner who suggested pursuing this topic long before ref. \cite{Hook:2017vyc} appeared. We also thank Csaba Cs\'aki, Anson Hook, Markus Luty, and Yuri Shirman for clarifying discussions.
This work was supported in part by the DOE under grant DE-SC-000999.

\end{document}